\documentclass[referee]{raa}
\pdfoutput=1
\usepackage{graphicx,times}
\usepackage{natbib}
\usepackage{amssymb,amsmath}
\usepackage{appendix}
\bibpunct{(}{)}{;}{a}{}{,}

\usepackage[a4paper=true,driverfallback=dvipdfm,pagebackref=true]{hyperref}
\hypersetup{colorlinks = true, linkcolor = green, anchorcolor = red, citecolor = blue, filecolor = red, urlcolor = red}

\begin{document}

   \title{The Feasibility and Flexibility of Selecting Quasars by Variability Using Ensemble Machine Learning Algorithms}

   \volnopage{Vol.0 (20xx) No.0, 000--000}      
   \setcounter{page}{1}          

   \author{Da-Ming Yang
      \inst{1,2}
   \and Zhang-Liang Xie
      \inst{1,2}
   \and Jun-Xian Wang
      \inst{1,2}
   }

   \institute{CAS Key Laboratory for Researches in Galaxies and Cosmology,
University of Science and Technology of China, Chinese Academy of Sciences, 
Hefei, Anhui 230026, China; {\it ydm2016@mail.ustc.edu.cn}\\
        \and
             School of Astronomy and Space Science, University of Science and Technology of China, Hefei 230026, China
\vs\no
   {\small Received~~20xx month day; accepted~~20xx~~month day}}

\abstract{ In this work we train three decision-tree based ensemble machine learning algorithms (Random Forest Classifier, Adaptive Boosting and Gradient Boosting Decision Tree respectively) to study quasar selection in the variable source catalog in SDSS Stripe 82. We build training and test samples (both containing 1:1 of quasars and stars) using the spectroscopic confirmed sources in SDSS DR14 (including 8330 quasars and 3966 stars).
We find that, trained with variation parameters alone, all three models can select quasars with similarly and remarkably high precision and completeness ($\sim$ 98.5\% and 97.5\%), even better than trained with SDSS colors alone ($\sim$ 97.2\% and 96.5\%), consistent with previous studies. Through applying the trained models on the variable sources without spectroscopic identifications, we estimate the spectroscopically confirmed quasar sample in Stripe 82 variable source catalog is $\sim$ 93\% complete (95\% for $m_i<19.0$). Using the Random Forest Classifier we derive the relative importance of the observational features utilized for classifications. We further show that even using one- or two-year time domain observations, variability-based quasar selection could still be highly efficient.
\keywords{quasars: general --- catalogs --- methods: data analysis}
}

   \authorrunning{D.-M. Yang, Z.-L. Xie \& J.-X. Wang}            
   \titlerunning{Selecting Quasars by Machine Learning}  

   \maketitle

%
%
\section{Introduction}\label{s_intro}

Quasars, as one of the most luminous celestial objects in the universe, are powered by accreting supermassive black holes (SMBHs) in galactic nuclei.
Building quasar catalogs is of great significance to research on SMBH accretion, galaxy evolution and cosmic structure. 
Since spectroscopic observations are always expensive in observing time, pre-selecting quasar candidates from large area photometric observations is an essential topic in this area.
Up to date, the largest samples of quasars were selected based on their optical/UV colors which are often different from those of stars \citep[e.g.][]{2002AJ....123.2945R, 2018A&A...613A..51P,2019ApJS..240....6Y}.
Infrared colors are also useful to select active galaxies and quasars \citep[e.g.][]{Stern2005,Lacy2007,Donley2012}.
Meanwhile, quasar candidates can be pre-selected based on multi-band detections, e.g., in X-ray, or in radio band.

In the era of time domain astronomy, selecting quasar candidates based on variation properties emerges as a new frontier. Flux variation in multi-band is one of the most prominent characteristics of quasars and active galaxies. In optical, quasars are often more variable than stars \citep{2007AJ....134.2236S}, and studies have shown that quasars can be efficiently selected based on their optical variation properties \citep[e.g.][]{2011ApJ...728...26M, 2014ApJ...782...37C}.

Variation-based quasar selection from large area time domain surveys could be uniquely helpful since it is free from the known biases suffered by the traditional color selection approach \citep[e.g.][]{2011AJ....141...93B,2019ApJS..242...10S}. For example, the traditional optical/UV color selection is insensitive to quasars at the redshift range of 2.2 to 3 in which their colors are similar to that of normal quasars \citep[e.g.][]{Schneider2010}.

While quasars can be selected using empirical cuts in the space of variation parameters \citep[e.g.][]{2011ApJ...728...26M,Schmidt2010,2011AJ....141...93B}, machine learning algorithms have been adopted by several studies to improve the precision and completeness of the selection. In these studies, various algorithms have been trained using various datasets.
\cite{2014MNRAS.439..703G} used Slepian wavelet variance (SMW), Damped Random Walk (DRW) and Structure Function (SF) at the same time to extract variability features from Catalina Real-time Transient Survey (CRTS) quasar samples, along with color features to train different ensemble learning classifiers, including Random Forest, Extremely randomized trees, AdaBoost and Gradient tree boosting. They also used SDSS Stripe 82 data to test their sample selection criteria.
\cite{2016ApJ...817...73H} utilized color and variability features and trained a Random Forest Classifier (RFC) to identify quasars on Pan-STARR1 3$\pi$ survey, in which each quasar was observed typically 7 times in each of its five bands within 3.5 years.
Similarly, \cite{2019ApJS..242...10S} used RFC to classify AGNs in QUEST--La Silla AGN variability survey. 
\cite{2018ApJ...869..178T} employed Support Vector Machine (SVM) to classify variable quasars in Sloan Digital Sky Survey Stripe 82, 
and \cite{2011ApJ...735...68K} also adopted SVM to identify variable quasars in MAssive Compact Halo Object (MACHO) database.

While these works have shown that the machine learning selection of quasars by variability is a promising approach, it would be useful to investigate how the performance of the machine learning approach depends on the availability of bands and the length of the light curves from which the variability parameters are measured. 
In this work we present an extensive study of machine learning selection of quasars by variability. 
We use the SDSS Stripe 82 variable source catalog to train three decision-tree based machine learning algorithms, including RFC, Adaptive Boosting \citep[AdaBoost,][]{Freund95adecision-theoretic} and Gradient Boosting Decision Tree \citep[GBDT,][]{FRIEDMAN2002367, Mason99boostingalgorithms}.
The ground-truth labels of objects we use in this work come from SDSS Data Release 14 (DR14) spectroscopic database. 
In \S\ref{s_data}, we introduce the dataset we used. 
In \S\ref{s_model}, we introduce the variability features and the training procedure, including the optimal hyper-parameters of the ensemble machine learning methods.
In \S\ref{s_performance}, we confirm that the three machine learning classifiers could yield remarkably high precision and completeness in classifying quasars. 
In \S\ref{s_completeness} we apply the trained algorithms to classify variable sources without spectroscopic identifications, and estimate the completeness of the quasar sample in SDSS Stripe 82.
We present and discuss the relative importance of the observational features (variability, color, from various bands) we utilized in \S\ref{s_importance}.
The effects of imbalanced samples (the ratios of quasars and stars used in training and test samples are not equal to 1) are discussed in \S\ref{s_imbalance}.
We explore the dependence of quasar classification on the length of light curves in \S\ref{s_oneyear}. 
Finally brief conclusions are given in \S\ref{s_conclusion}.

\section{SDSS Stripe 82 variable sources} \label{s_data}

SDSS Stripe 82 is a 290 deg$^2$ equatorial field, which has been repeatedly scanned $\sim$ 60 times in $ugriz$ within $\sim$ 10 years by the Sloan Digital Sky Survey \citep{2007AJ....134..973I}.  
A catalog of 67,507 variable sources in SDSS Stripe 82 was built by \cite{2007AJ....134..973I}. 
Thanks to the long duration and the large number of visits for each source, variation parameters could be measured with considerably high accuracy.  
The catalog, together with the SDSS spectroscopical identifications,  is uniquely useful to promote systematical study and understanding of variation-based quasar classification.
Such study could be essential to quasar selection from upcoming large area time domain surveys (e.g. LSST).

This catalog was built following the criteria listed below:
\begin{itemize}
\item {unresolved in imaging data, with photometric error below 0.05 mag in at least one band}
\item {processing flags BRIGHT, SATUR, BLENDED, or EDGE are not set in any band}
\item {at least 10 photometric observations in the ${g}$ and ${r}$ bands}
\item {the median $g$-band magnitude brighter than 20.5}
\item {root-mean-square scatter $>$ 0.05 magnitude and $\chi^2$ per degree of freedom larger than three in both g and r bands, which mean the variation is statistically significant (see the discussions in \citealt{2007AJ....134.2236S} for the details)}
\end{itemize}

We match the catalog with a matching radius of 2\arcsec\ with SDSS Data Release 14 (DR14, \citealt{2018ApJS..235...42A}) for spectroscopical identifications. Among the variable sources,  8330 are identified as quasars and 3966  as stars. The rest 48716 sources are left unlabeled.
The spectroscopically identified sources are utilized to train and test our machine learning classifiers, which are also utilized to classify those unlabeled sources. 

\section{Observational Features and Machine Learning Models} \label{s_model}

\subsection{Observational Features} \label{s_features}

We use a damped random walk (DRW; Also named as Ornstein-Uhlenbeck process) process to fit the light curves. The DRW process is a stochastic process defined by an exponential covariance matrix between $t_i$ and $t_j$:

\begin{equation} \label{e_drw}
	S_{DRW}(\Delta_{t}) = \sigma^2 exp(-|\frac{\Delta_{t}}{\tau}|)
\end{equation}

where $\sigma$ is the long-term deviation of variability, and $\tau$ the characteristic time scale of DRW. It is essentially a random walk with a self-correcting term that pushes any deviations back toward the mean value of the random walk itself.
Various studies have shown that the DRW process (also named Ornstein-Uhlenbeck process)) could well describe the observed UV/optical variations of active galaxies and quasars \citep[e.g.][]{2009ApJ...698..895K,2010ApJ...708..927K,2010ApJ...721.1014M}
\footnote{Though deviations from DRW at extremely short and extremely long timescales have been reported (e.g \cite{2011ApJ...743L..12M}, \cite{2013ApJ...765..106Z}, \cite{2017ApJ...847..132G}).
More sophisticated models include mixture of Ornstein-Uhlenbech (OU) processes \citep{2011ApJ...730...52K}, continuous auto regression and moving average (CARMA) model \citep{2014ApJ...788...33K, 2016A&A...585A.129S, 2017MNRAS.470.3027K}, and a broken powerlaw shaped power density spectrum \citep[e.g.][]{2016ApJ...825...56Z} have been introduced}.
By now, comparing with many other deterministic and stochastic models, the DRW process is the best model for SDSS Stripe 82 quasar light curves \citep{2013A&A...554A.137A}. Meanwhile, the DRW parameters had been found useful to distinguish quasars from variable stars. For instance,  \cite{2011ApJ...728...26M} found that the distribution of $\tau$ (in the observed frame) peaks around 500 days for quasars in Stripe 82,  but $\sim$ 1 day for other objects, showing selecting quasars out of variable sources is highly promising even without color information.

In this work we use the DRW parameters as the input observational features to train and test our machine learning classifiers.
We use the software JAVELIN \citep{2011ApJ...735...80Z, 2013ApJ...765..106Z} to fit each SDSS Stripe 82 light curve to measure $\tau$ and  $\sigma$.
A  total of 10 DRW parameters (for 5 SDSS bands) were obtained for each variable source.
The DRW fitting failed for 2 stars and 110 unlabelled sources, which are excluded from further analyses. 
We also note some stars with ``peculiar" fitted DRW parameters, which however do not affect the analyses in this work. 
An example light curve of such sources is presented In Appendix \ref{a_star}, which results in an unreasonably large $\sigma$.
We presume that the DRW fitting may fail or yield unreasonable results is owing to the fact that the variability of such sources  is far from a DRW process.

For light curves with small number of epochs, the DRW fitting might yield parameters with huge uncertainties. Therefore, we also measure the maximum variation amplitudes of each source (in $gri$ band, which have considerably better SDSS photometry comparing with $u$ and $z$), and include them as input to the classifiers. The relative importance of these input features are presented and discussed in \S\ref{s_importance}.

It would be interesting to examine, for the variation selected sources in this study,  whether the variation features alone can better classify quasars comparing with colors, and whether combining variation and color features can further improve the classification. The color features of the variable sources ($u-g$, $g-r$, $r-i$ and $i-z$) are thus also considered, which were calculated using the median photometry in each band.

\subsection{Machine Learning Models}\label{s_ml}

\begin{figure*}[htp]
    \centering 
    \includegraphics[width=0.76\linewidth]{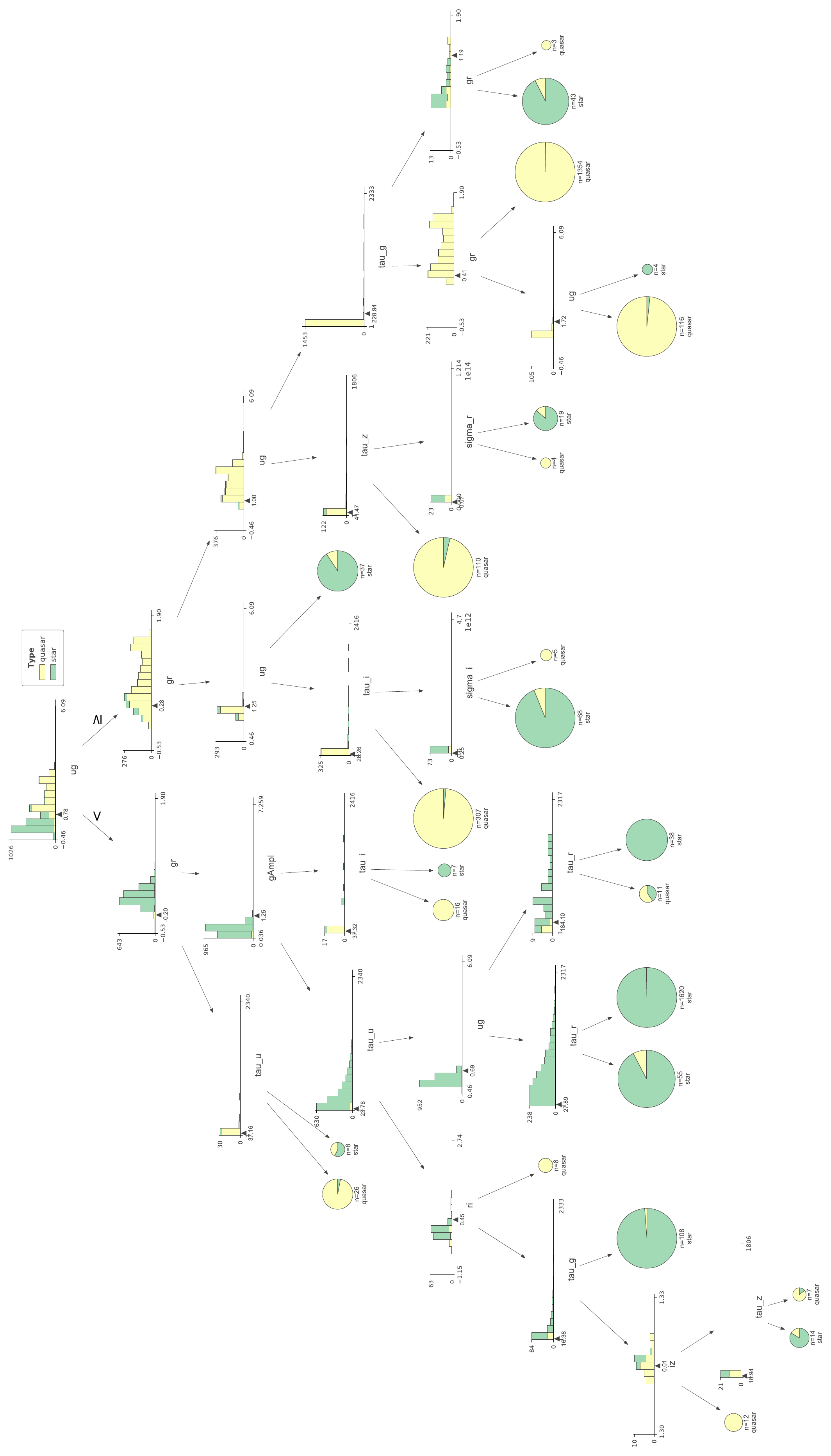}
    \caption{The tree-like structure of an example decision tree in this work (pruned for better visualization). Visualization is done with python package \textbf{dtreeviz}.}
    \label{f_dt}
\end{figure*}

We use RFC as our major classifier in this study. RFC is one of the most popular and powerful supervised machine learning methods. It belongs to a subclass called ensemble learning method,  which includes many weak classifiers, and all the weak classifiers collaboratively make a final decision. In the case of random forest, it includes hundreds (or even thousands) of decision trees, each of them is trained by the training set. An example of decision tree trained in this study is presented in Fig. \ref{f_dt}. 

Each internal node stands for a selection rule on a specific feature, which splits the node into two branches. Each leaf node (nodes at the bottom of a tree) stands for a class, in our case, a quasar or a star. In real training, a branch of a decision tree will stop growing deeper once the purity of the newest node reaches the preset value (not necessary to be $100\%$) , and the node will become a leaf node. 

We also adopt two other decision-tree-based models, namely, Adaptive Boosting \citep[AdaBoost,][]{Freund95adecision-theoretic} and Gradient Boosting Decision Tree \citep[GBDT,][]{FRIEDMAN2002367, Mason99boostingalgorithms}. Unlike random forest, they are models using boosting but not bagging strategy (random forest is essentially a bagging ensemble learning model). AdaBoost is adaptive because the misclassified samples will be used specifically in the next iteration, and the outputs of decision trees (``weak learners") in each epoch are combined into a weighted sum as a final output. 
For GBDT, in each iteration the algorithm will try to diminish the value of loss function.
Both AdaBoost and GBDT are consecutive ensemble models.
Contrarily, random forest, by concept, is a parallel ensemble model,  and its purpose is to build numerous independent weak classifier and average their results. 
They all belong to ensemble learning methods,  which help improving machine learning results by combining several base models to produce an optimal predictive model. Generally ensemble learnings tend to have better performances and are less likely to overfit.

The machine learning frame we use is scikit-learn (\cite{scikit-learn}; formerly scikits.learn). All models require hyperparameters setting which is independent of the dataset. By now, hyperparameters optimization is mostly implemented empirically. However, a practical approach without much experience is Grid Search. We present in Table \ref{t_hp} our best hyperparameters derived with sklearn.model\_selection.GridSearchCV below for each model (other parameters not mentioned are left as default).

\begin{table}[ht]
    \center
    \begin{tabular}{c | c c}
    \hline
    \hline
    \textbf{model} & \textbf{hyperparameter} & \textbf{value} \\
    \hline
    
    \textbf{RFC}
    
    & \textsf{n\_estimators} &  500\\
     
    & \textsf{oob\_score} & \emph{True} \\
    
    \hline
    \textbf{AdaBoost}
    
    & \textsf{n\_estimators} &  500\\
    
    & \textsf{learning\_rate} &  0.08\\
    
    & \textsf{algorithm} & \emph{SAMME} \\
    
    & \textsf{max\_depth} & 6 \\
    
    & \textsf{min\_samples\_split} & 20 \\
    
    & \textsf{min\_samples\_leaf} &5 \\

    \hline 
    \textbf{GBDT}
    
    & \textsf{n\_estimators} &  500\\
    
    & \textsf{learning\_rate} &  0.1\\
    \hline
    \hline

    \end{tabular}
    \caption{Hyperparameters setting for all models}
    \label{t_hp}
\end{table}

Among the spectroscopically identified sources, we randomly select 3000 quasars and 3000 stars to train our models. 
A sample of 600 quasars and 600 stars (an 1:1 sample), mutually different from the training sample, is then built to test the trained classifiers. 
Two indices, namely precision and recall (see equations \ref{e_pr}), are used to evaluate the performance of a trained model. 

\begin{equation}\label{e_pr}
\begin{aligned}
Precision_{star} &= \frac{\# \ of \ predicted \ true \ star}{\# \ of \ predicted \ star} \\ 
Recall_{star} &= \frac{\# \ of \ predicted \ true \ star}{\# \ of \ confirmed \ star} \\
Precision_{quasar} &= \frac{\# \ of \ predicted \ true \ quasar}{\# \ of \ predicted \ quasar} \\
Recall_{quasar} &= \frac{\# \ of \ predicted \ true \ quasar}{\# \ of \ confirmed \ quasar} \\
\end{aligned}
\end{equation}

where precision stands for the fraction of that a certain type of classifications is true, and recall the completeness of correct classification for a given class of objects. We repeat above random selections of training and test samples 100 times, and present the averaged performance in \S\ref{s_performance}. Note we adopt a ratio of 1:1 of stars versus quasars for both the training and test datasets. While this is a common approach in machine learning studies, we discuss in \S\ref{s_imbalance} the effects of imbalanced datasets.

\section{Results and Discussion}\label{s_result}
\subsection{The high performance of the machine learning algorithms}\label{s_performance}

\begin{table*}[t]
    \center
    \begin{tabular}{c | c | c c c}
    \hline
    \hline
    \textbf{feature\ mode} & \emph{} &\textbf{RFC} &\textbf{AdaBoost} &\textbf{GBDT} \\
    
    \hline
    
    \textbf{all\ features}
    
    & \emph{$P_{quasar}$} & $99.00\pm0.04\%$ & $99.05\pm0.04\%$ & $98.90\pm0.04\%$\\

    & \emph{$R_{quasar}$} & $98.84\pm0.04\%$ & $99.02\pm0.04\%$ & $98.85\pm0.04\%$\\

    & \emph{$P_{star}$} & $98.84\pm0.04\%$ & $99.02\pm0.04\%$ & $98.85\pm0.04\%$\\

    & \emph{$R_{star}$} & $99.00\pm0.04\%$ & $99.05\pm0.04\%$ & $98.90\pm0.04\%$\\
    \hline

    \textbf{color\ features}
    
    & \emph{$P_{quasar}$} & $97.16\pm0.06\%$ & $97.20\pm0.06\%$ & $97.23\pm0.06\%$\\
    
    & \emph{$R_{quasar}$} & $96.64\pm0.07\%$ & $96.56\pm0.07\%$ & $96.43\pm0.07\%$\\
      
    & \emph{$P_{star}$} & $96.66\pm0.06\%$ & $96.59\pm0.06\%$ & $96.46\pm0.07\%$\\
          
    & \emph{$R_{star}$} & $97.18\pm0.06\%$ & $97.22\pm0.07\%$ & $97.24\pm0.06\%$\\
    \hline
   
    \textbf{variability\ features}
    
    & \emph{$P_{quasar}$} & $98.56\pm0.04\%$ & $98.43\pm0.04\%$ & $98.31\pm0.05\%$\\
    
    & \emph{$R_{quasar}$} & $97.50\pm0.06\%$ & $97.81\pm0.05\%$ & $97.59\pm0.06\%$\\
      
    & \emph{$P_{star}$} & $97.53\pm0.06\%$ & $97.82\pm0.05\%$ & $97.61\pm0.06\%$\\
          
    & \emph{$R_{star}$} & $98.57\pm0.04\%$ & $98.44\pm0.04\%$ & $98.32\pm0.05\%$\\
    \hline
    \hline
    
    \end{tabular}
    \caption{The performance of three machine learning algorithms. The mean values (of precision and recall) and the corresponding errors of the mean values are derived through averaging results from 100 trials. 
In the  ``all features" model both color and variability features (see \S\ref{s_features}) are utilized. 
   }
    \label{t_result}
\end{table*}

\begin{figure}[ht]
    \centering 
    \includegraphics[width=1\linewidth]{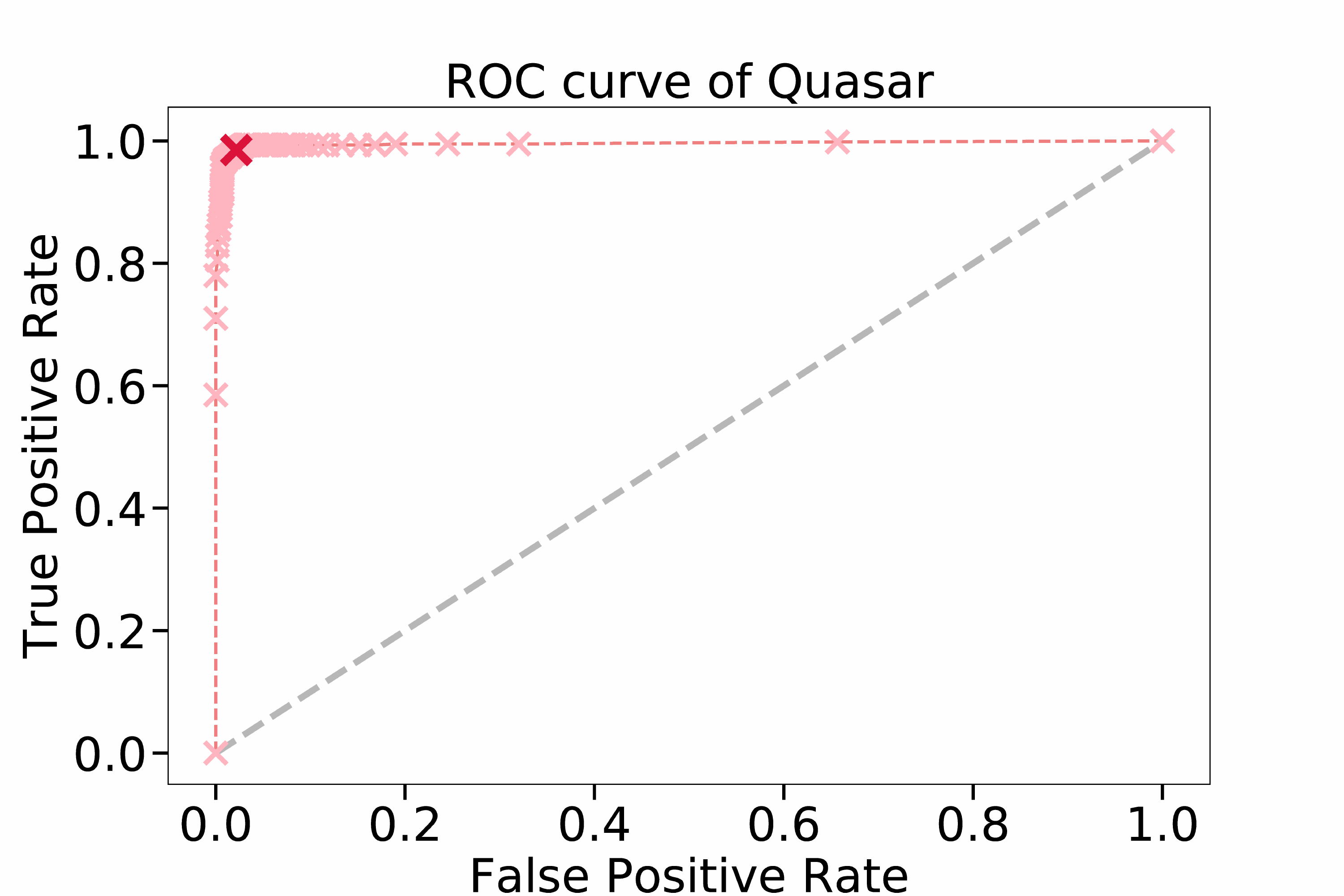}
    \caption{ROC curve of quasar (i.e., ``true sample" here means quasar) given by the RFC classifier. The red cross in the plot is the best threshold 0.42, the nearest point to (0.0,1.0).}
    \label{f_roc}
\end{figure}

In Table \ref{t_result},  we present the performance of the three classifiers evaluated with the test samples. 
The values in the table are obtained through averaging 100 trials (hereafter the same), and the errors of the mean values are also presented.
We find similarly high performances obtained with all three machines, showing each of them 
is competent enough for such a study. In Fig. \ref{f_roc}, we present the receiver operating characteristic (ROC) curve yielded by the RFC classifier, created by plotting the true positive rate (TPR) against the false positive rate (FPR) at various thresholds. The ROC curve appears ideal for a machine learning task, indicating the distributions of the two types of objects are well separated.

We also find that using the variability features alone can yield better performances comparing with using only color features. 
This  clearly demonstrates the high efficiency of selecting quasars with time domain observations. Putting variability and color features together would further enhance 
the selection accuracy, with $\sim$ 99.0\% precision and recall achieved for all three machines.

Next, we briefly compare our results with previous representative works on quasar selection out of SDSS stripe 82 variable sources.

\cite{2011ApJ...728...26M} selected 10,024 SDSS Stripe 82 variable sources with $\emph{i} < 19.0$. Among them 1,490 ($\sim$ 15\%) are spectroscopically confirmed quasars, and the rest were considered as non-quasars. 
They found that simple cuts in DRW parameters (such as $\tau$ $>$ 100 days), aided with the color selection, could achieved a precision of $93.8\%$ (named as efficiency in the paper) and recall of $98.0\%$ (named as completeness) in selecting quasars
\footnote{Note the performance were estimated with the same dataset they used to define the thresholds thus could have been over-estimated.}. 
After corrected for the imbalance (see \S\ref{s_imbalance}), their precision and recall are 99.3\% and 93\% respectively. It can be seen that the analytical approach of \cite{2011ApJ...728...26M} utilized rather strict thresholds which could achieve high precision, but suffer considerable incompleteness.

\cite{2018ApJ...869..178T} used SDSS Stripe 82 variable sources catalog to train a supportive vector machine (SVM). They constructed a dataset which consists of 7,714 confirmed quasars and 2,141 stars. They used various sets of variability features to train the SVM. Their test dataset contains 1000 quasars and 1000 stars, and the rest of the sources were used to train the machine. 
Using DRW parameters measured with JAVELIN, they obtained averaged precision and recall of 93.8\% and 98.6\%
\footnote{Private communications with the authors show there are mistakes with their definitions of precision and recall in their original published paper, and an erratum is to be submitted. The values we quoted here are corrected ones given by the authors.}.
While their recall is similar to our results, their precision is considerably lower. This could be partly due to the fact that they used smaller sample of 1141 stars to train the classifier. Also, SVM is a weak classifier, like a single decision tree in the random forest model. For comparison, the precision and recall of quasar given by a single decision tree are $\sim98.0\%$ and $\sim97.0\%$. Meanwhile, their precision is considerably lower than their recall, mainly because of the imbalanced test sets.

\cite{2014MNRAS.439..703G} used Slepian wavelet variance, Damped Random Walk and Structure Function at the same time to describe the variability features. Using variability alone, they got 96.5\% recall (named as completeness in the literature) and 95.0\% precision (named as purity in the literature) for their RFC. With the aid of color features, they got 99.3\% recall and 99.0\% precision. 
Using DRW parameters alone, we achieve similar recall and precision, confirming the results of \cite{2014MNRAS.439..703G}.

\subsection{The completeness of the spectroscopically confirmed quasar sample in Stripe 82}\label{s_completeness}

\begin{figure}[ht]
    \centering 
    \includegraphics[width=1\linewidth]{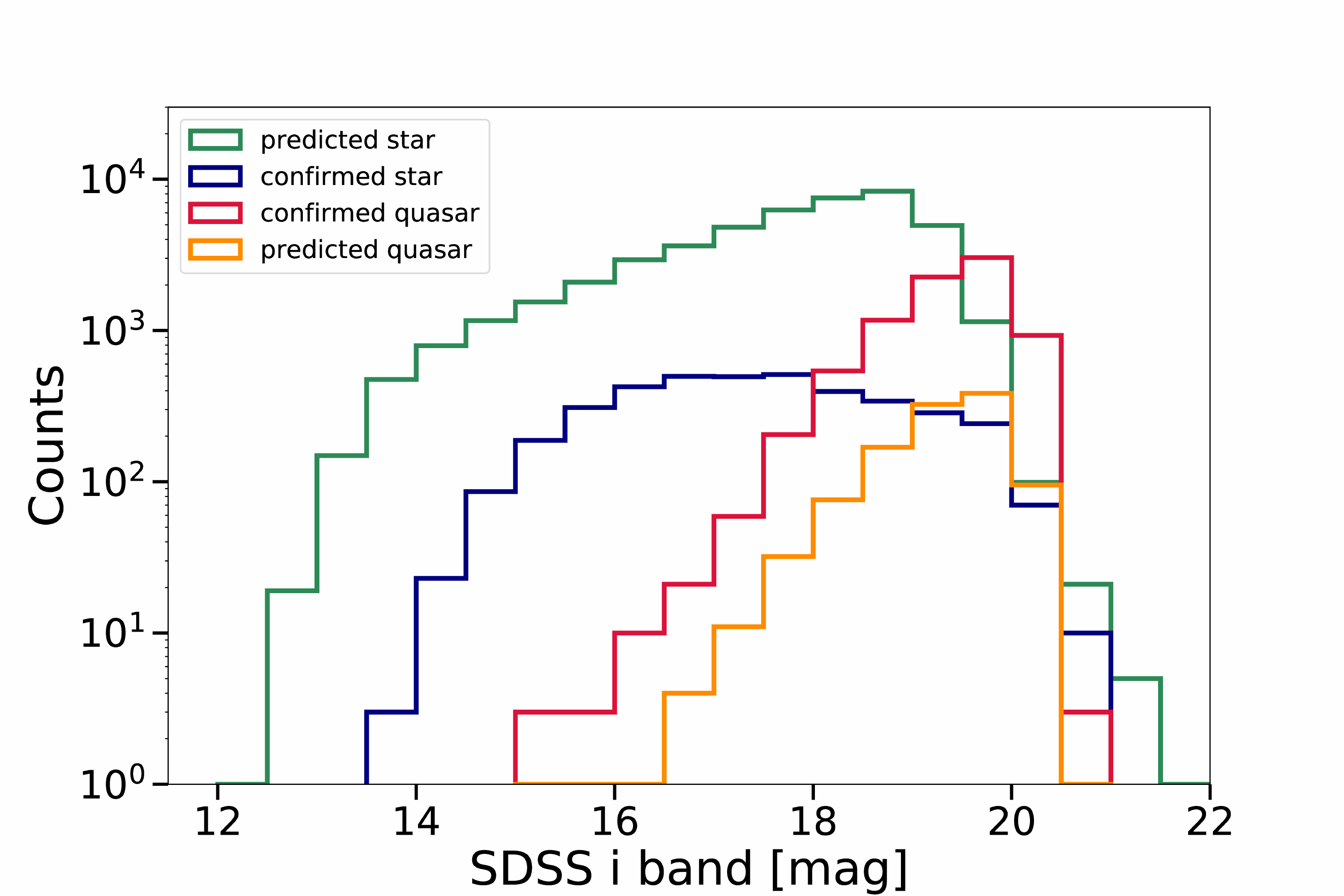}
    \caption{SDSS  i-band magnitude distributions of spectroscopically confirmed and RFC predicted quasars and stars. Predicted sources are those without spectroscopical identifications, but classified by our random forest model.}
    \label{f_iband}
\end{figure}

We then apply our trained models to select potential quasars out of the unlabeled sources in the Stripe 82 Variable Source Catalog. With RFC, we classify 1,105 of them as predicted quasars, and 47,501 as stars (As mentioned in \S\ref{s_features}, the DRW fitting failed for 110 unlabelled sources, which are most likely to be stars, thus having no influence on further discussion on completeness). Similar numbers are obtained using AdaBoost and GBDT. We plot the i-band magnitude distributions of the RFC predicted quasars and stars in Fig. \ref{f_iband}, together with those of the confirmed ones. While confirmed quasars and stars have significantly  different magnitude distributions,  it's interesting to note that the i-band magnitude distributions of predicted and spectroscopically confirmed quasars are similar, and so do those of stars. 

Recalls, i.e., $\frac{\#\ of\ true\  predicted\ *}{\#\ of\ real\ *}$ (*: quasar / star), do not change with the absolute values of the numbers of quasar and star in the samples, whereas precisions, i.e., $\frac{\#\ of\ true\  predicted\ *}{\#\ of\ predicted\ *}$, may be affected greatly due to the imbalance, as mentioned in \S\ref{s_imbalance}. Assuming recalls of the unlabeled sources are the same as those in Table \ref{t_result}, we can estimate the numbers of real sources with equations \ref{e_predict}, which are derived from the definitions of recalls in equations \ref{e_pr}.
After correcting the incompleteness and contaminations due to misclassifications, we finally expect there are $\sim$ 633 real quasars among the 48,716 unlabeled sources, $\sim633\times(1-0.9984)\sim7$ of them could have been misclassified as stars, and the sample of the 1,105 predicted quasars has an precision of $\sim\frac{633-7}{1105}\times100\%\sim57\%$.
This suggests the spectroscopically confirmed sample of quasars among the variable sources has a completeness of $\sim\frac{8330}{8330+633}\times100\%\sim93.0\%$, similar to the estimated completeness of SDSS quasars  from small spectroscopically complete samples \citep[$>$90\%,][]{2002AJ....123.2945R,2002AJ....124.2364I,2005AJ....129.2047V,2015ApJ...811...95P}.

\begin{equation}\label{e_predict}
\begin{aligned}
(1 - Recall_{star}) \times &\# \ of \ real \ star =  \\ &\# \ of \ predicted \ quasar - Recall_{quasar} \times \# \ of \ real \ quasar \\
(1 - Recall_{quasar}) \times &\# \ of \ real \ quasar =  \\ &\# \ of \ predicted \ star - Recall_{star} \times \# \ of \ real \ star \\
\end{aligned}
\end{equation}

\begin{figure}[ht]
    \centering 
    \includegraphics[width=1\linewidth]{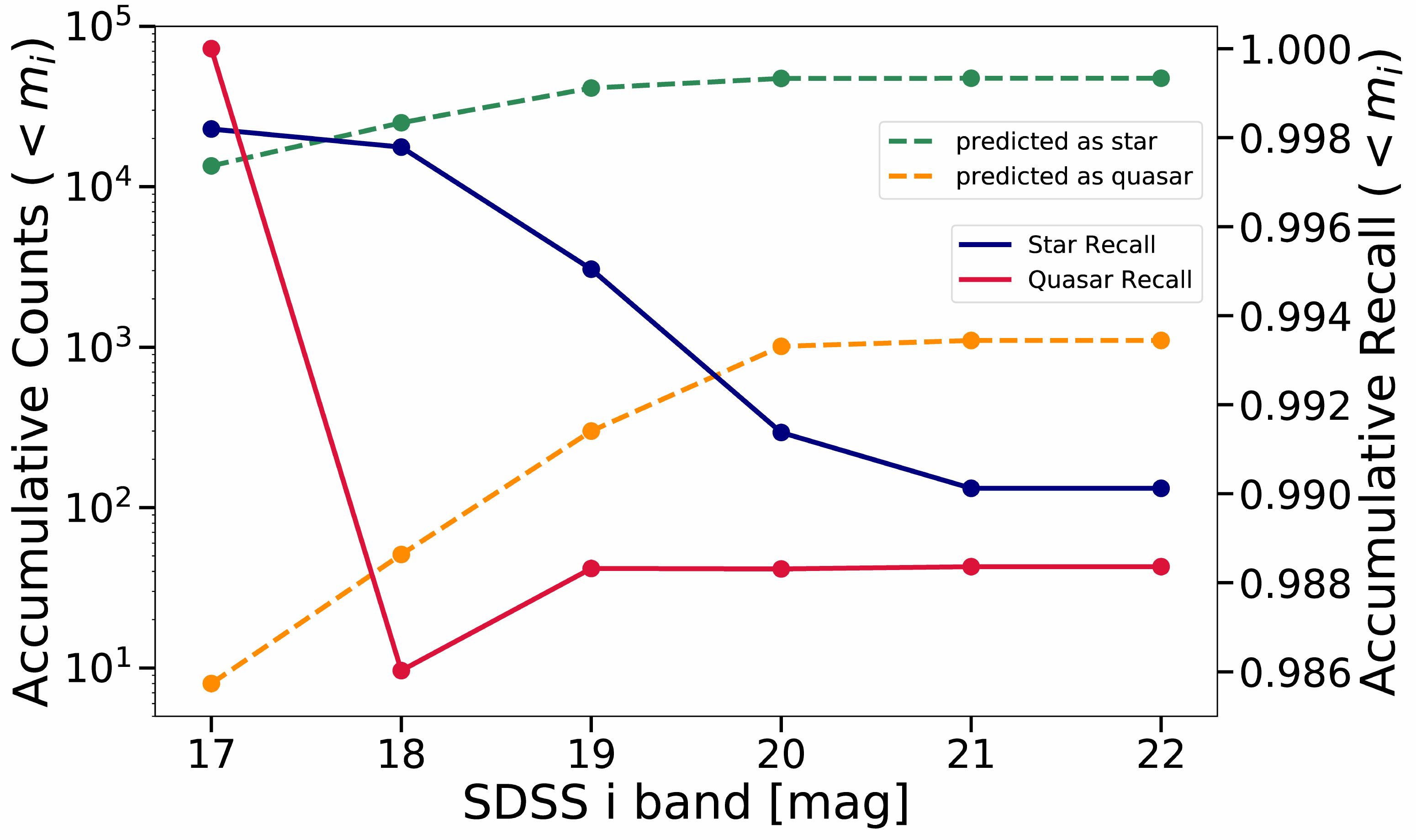}
    \caption{The RFC performance (accumulative recall, averaged through 100 testing runs), and the numbers of predicted quasars/stars (out of the unlabeled sources), as a function of limiting magnitude m$_i$.}
    \label{f_mbin}
\end{figure}

\begin{figure}[hbpt]
    \centering 
    \includegraphics[width=1\linewidth]{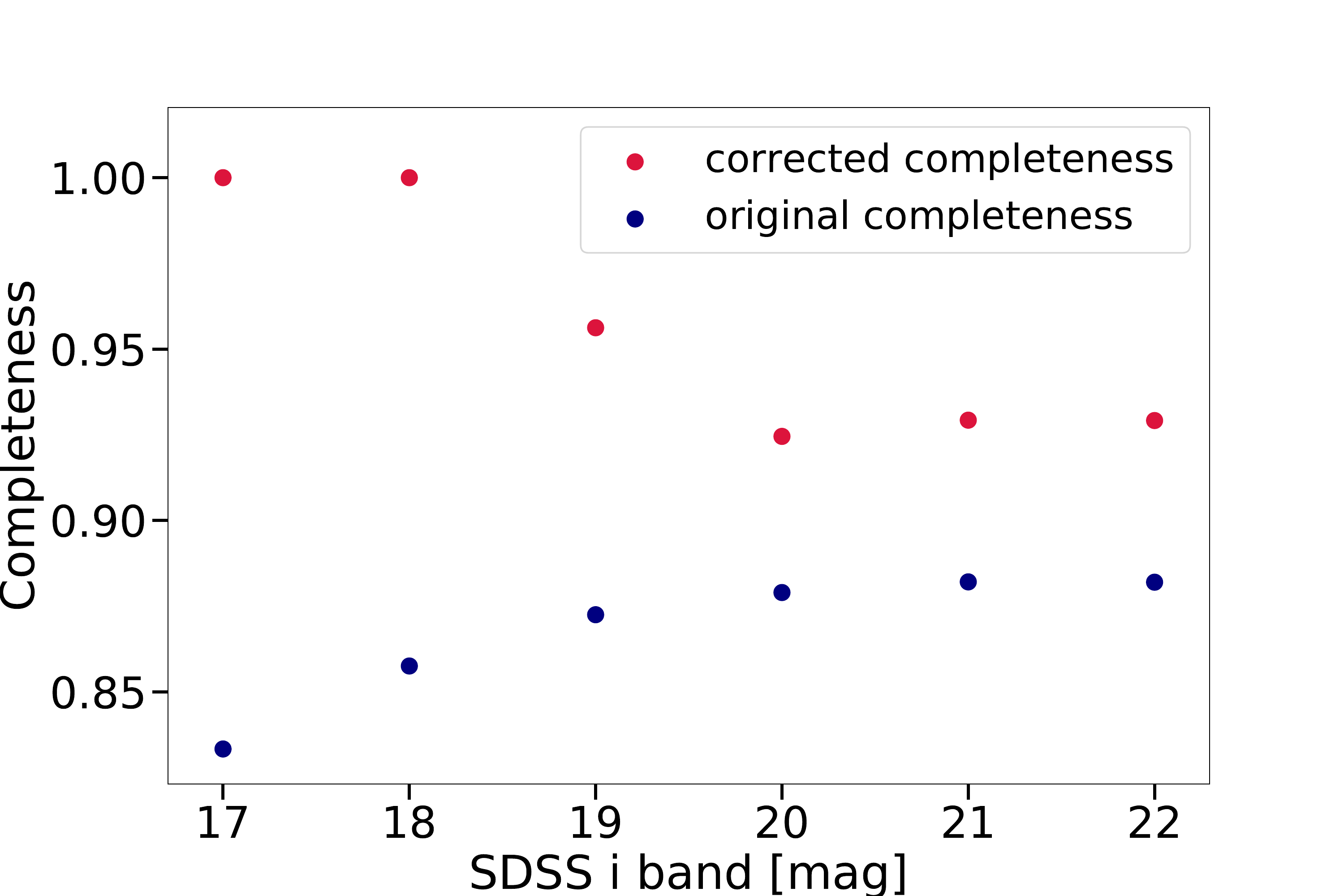}
    \caption{The estimated completeness of spectroscopically confirmed quasars in the SDSS Stripe 82 variable source catalog, as a function of limiting magnitude m$_i$. The ``original completeness" is calculated simply using the numbers of predicted quasars from our RFC classifier, i.e., (spectroscopically~confirmed)/(confirmed + predicted).}
    \label{f_comp}
\end{figure}

We see from Fig. \ref{f_iband} that, among the variables sources, stars significantly outnumber quasars at brighter magnitudes, and at the faint end quasar is the dominant population. 
We thus expect that at brighter magnitudes, the predicted quasar sample suffers stronger contaminations from misclassified stars. Such effect could be corrected through repeating 
the calculations describe in the above paragraph, but at different limiting magnitudes.

In Fig. \ref{f_mbin} we plot the recalls of quasars and stars we measured with test samples, as a function of limiting $i$ band magnitude. 
Utilizing the measured recall, and the number of predicted quasars and stars, we calculate the corrected completeness (as a function of limiting magnitude, i.e. $<m_i$)  in Fig. \ref{f_comp}. This indicates that SDSS quasar sample is highly complete ($\sim$ 95\%) at $i$ $<$ 19, which is also consistent with the estimated completeness based on small spectroscopically complete samples (\citealt{2005AJ....129.2047V}, $94.9\%$ for $i<19.1$; and \citealt{2015ApJ...811...95P}, $94.7\%$ for $i<19.1$).

The ``original completeness", calculated simply using the numbers of predicted quasars from our RFC classifier, is also plotted in Fig. \ref{f_comp} for comparison.
Note such completeness is significantly contaminated by stars which have been misclassified as quasars, and such contamination is stronger at brighter magnitudes as there are relatively more 
bright stars than quasars in the variable source catalog (see Fig. \ref{f_iband}). Such effect could explain the  even smaller  ``original completeness" at brighter magnitudes in Fig. \ref{f_comp}.

\subsection{Feature Importance}\label{s_importance}

Not all input features to the classifiers are equally useful in distinguishing quasars from stars. 
It's helpful to examine the relative importance of various features, particularly considering that
the available features are practically often limited by observational resources. 

\begin{figure}[htbp]
    \begin{center}
    \includegraphics[width=0.9\linewidth]{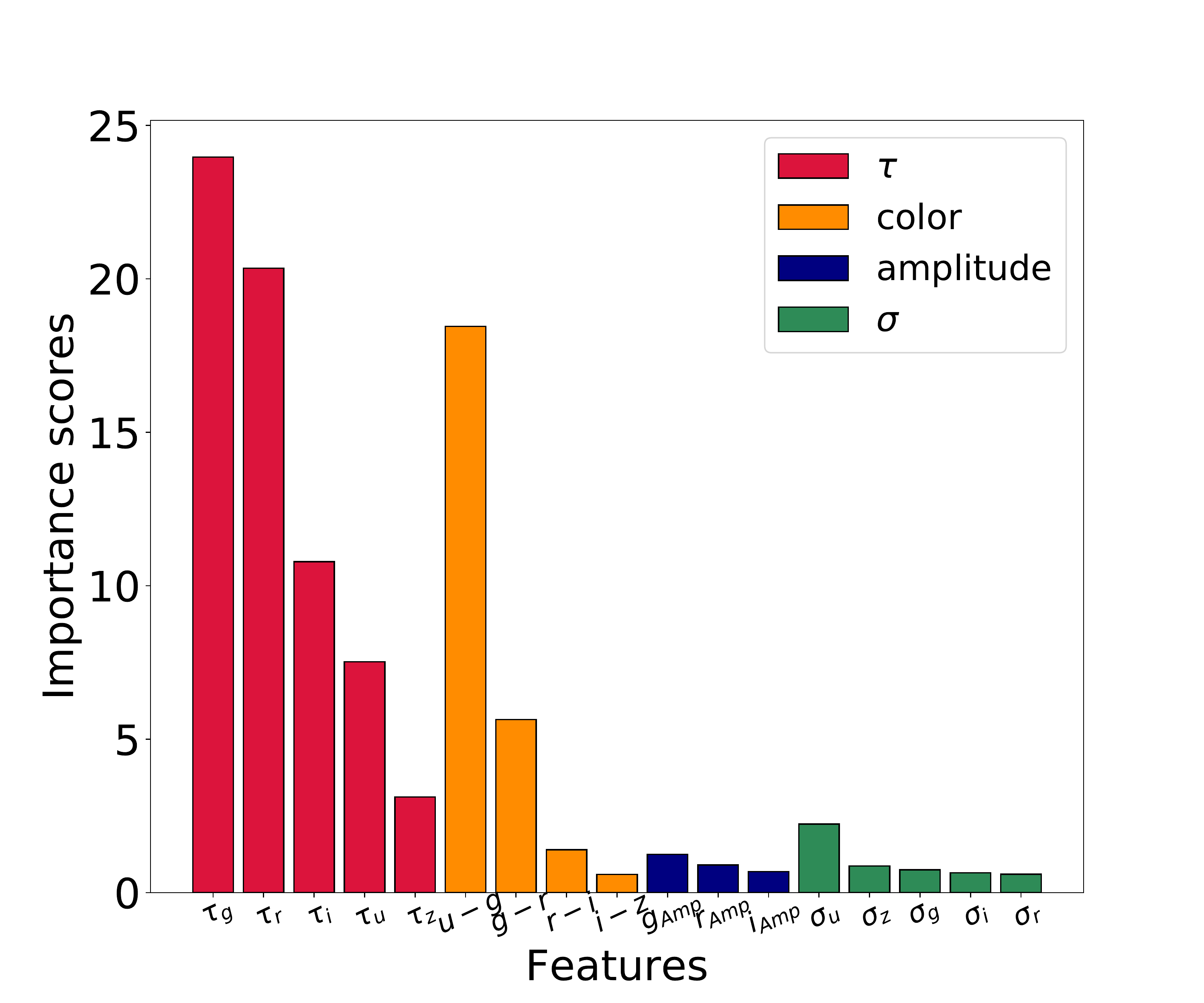}
    \caption{Ranking of feature importance (gini importance) given by RFC. Red (green) bars represent DRW parameter $\tau$ ($\sigma$) in five bands, yellow bars the maximum variation amplitudes in $gri$, and  blue bars the color features.}
    \label{f_importance}
    \end{center}
\end{figure}

A decision tree or random forest can generate features rankings by calculating so called "gini importance" or "mean decrease impurity" \citep{breiman1984classification}, which calculates each feature importance as the sum over the number of splits (across all tress in a random forest) that include the feature, proportionally to the number of samples it splits. In another word, features that can separate a larger set of samples into two pure enough subset have larger feature importances. We run 100 trials to get average scores of every features. We present the results in Fig. \ref{f_importance}, where significant diversity in the feature scores is seen.

\begin{figure}[htbp]
    \begin{center}
    \includegraphics[width=0.9\linewidth]{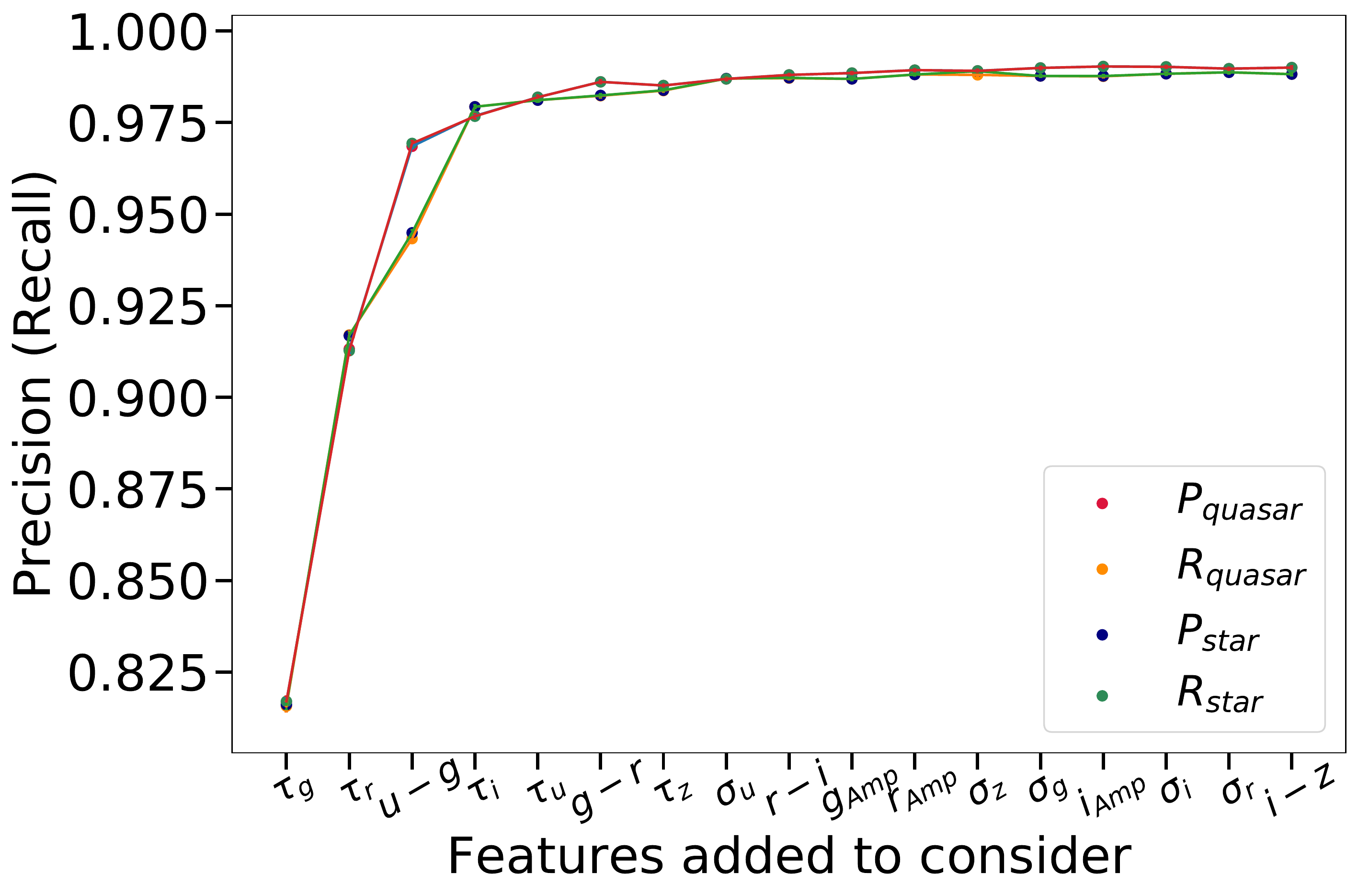}
    \caption{$Precisions$ and $recalls$ of different input feature sets. Features are added one at a time by the ranking of importance score (from left to right). Only $\tau_g$ is used in the first point, and all features are included in the last point. We notice a sharp rise in the first three points, for they represent the top three most importance features.}
    \label{f_allf}
    \end{center}

\end{figure}

We then add feature one at a time into the feature set in the order of their importance scores, to train and test the RFC model. 
The output $precision$ and $recall$ are plotted in Fig. \ref{f_allf}, where we clearly see that the performance of the classifier is dominantly driven by the first few features.

\begin{figure}[htbp]
    \begin{center}
    \includegraphics[width=0.9\linewidth]{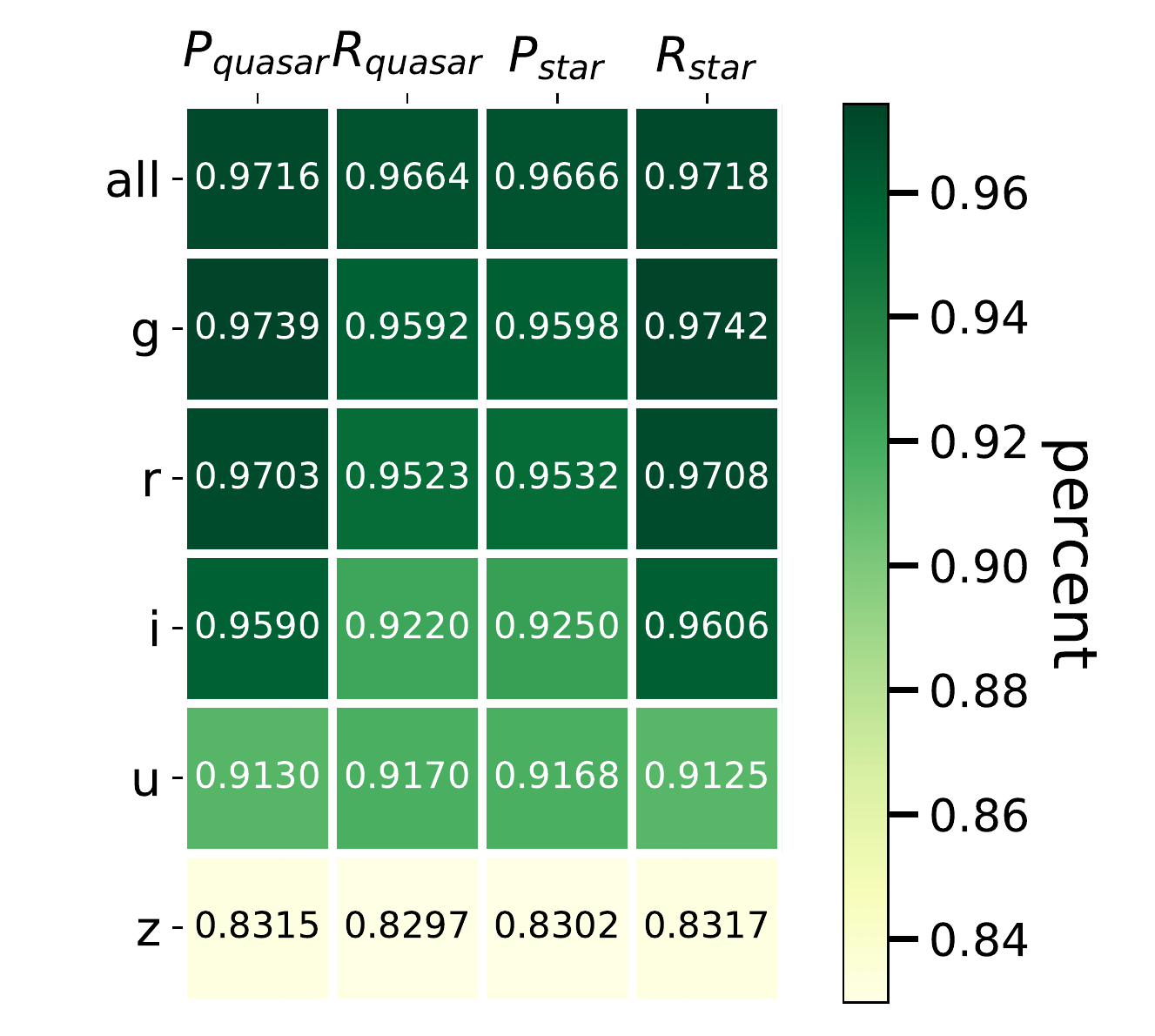}
    \caption{The achieved performance of RFC of using variation features from one band only, 
    e.g., a block marked by \emph{g} represents performances that only variation features of \emph{g-band} are used, which are $\tau_g$, $g_{Amp}$ and $\sigma_{g}.$}
    \label{f_band}
    \end{center}
\end{figure}

In Fig. \ref{f_band} we also plot the measured $precisions$ and $recalls$ by using variation features in one band only. We see that $g$ and $r$ bands have similarly good performance, likely because SDSS quasars have the best photometry in $g$ and $r$. Meanwhile the performance in $z$ band is the worst, which could be attributed to its much worse photometry  and the fact that quasar variations are much weaker at longer wavelength.
Note that the performance using only $g$ or $r$ band variation features alone is already comparable to that using all SDSS colors.

\subsection{Effects of Imbalanced Samples} \label{s_imbalance}

We previously present and discuss the efficiencies of the machine learning models using training and test samples with 1:1 ratio of quasars and stars. 
However actual datasets are often heavily imbalanced. For instance, among the SDSS Stripe 82 variable sources we used in this work, there are 8330 spectroscopically confirmed quasars, 
but only 3966 stars. More significantly, among the 48716 unlabeled sources, we only expect $\sim$ 630 quasars (see \S\ref{s_completeness}). 
Below we discuss the effects of sample imbalance in both the test sample and the training sample.

The effect of imbalance in the test sample (or the to-be-classified sample) is straightforward.
Comparing with an 1:1 sample, an imbalanced test sample would in principle yield identical $recall$ but different $precision$ for each type of the objects.
This is because, for instance, whether a quasar in the test sample could be correctly classified ($recall$) depends on the observed features of the quasar and how the classifier was trained, but has nothing to do with other sources in the test sample. Contrarily, the $precision$ of the quasar classifications do depend on the number of stars which have been mis-classified as quasars, thus the number of stars in the sample.

Let $\eta$ be the ratio of stars to quasars in the test sample. The expected precisions of quasars and stars from an imbalanced test sample can be expressed as:

\begin{equation}
\begin{aligned}
Precision_{quasar}=\frac{Recall_{quasar}}{Recall_{quasar}+\eta\times(1-Recall_{star})}\\
Precision_{star}=\frac{Recall_{star}}{Recall_{star}+1/\eta\times(1-Recall_{quasar})}
\label{e_imbalance}
\end{aligned}
\end{equation}

We can clearly see from the above equation that, a test sample with stars more than quasars ($\eta$ $>$ 1) would yield lower quasar precision (and higher star precision) comparing with the 1:1 sample (see also \S\ref{s_completeness}).

The effect of imbalanced training sample is more complicated and there is no simple analytical equation. 
In principle, if there are more stars than quasars in the training set, the machine learning model will likely learn more information about the stars. In this way, a star will be less likely misclassified as a quasar. But apparently, the shortage of this approach is that a quasar will be more likely misclassified as a star. This compromise shows the well known trade between precision and recall. 
Such imbalance is a common issue in the machine learning field, and many algorithms and methods have been brought forward to deal with it \citep[e.g.][]{5128907,Chawla:2004:ESI:1007730.1007733}. 
Generally, utilizing special sampling methods (e.g. resampling, over-sampling, under sampling) and adjusting loss function (e.g. cost-sensitive) are the most common ways to deal with the problem.

\subsection{Dependence on the length of light curves}\label{s_oneyear}

\begin{table*}[t]
    \center
    \begin{tabular}{c | c | c c c}
    \hline
    \hline
    \textbf{feature\ mode} & \emph{} &\textbf{one-year} &\textbf{two-year} &\textbf{ten-year} \\
    
    \hline

    \textbf{all\ features}
    
    & \emph{$P_{quasar}$} & 98.18 $\pm$ 0.05 \% & 98.29 $\pm$ 0.05 \% & 98.91 $\pm$ 0.04 \%\\

    & \emph{$R_{quasar}$} & 98.09 $\pm$ 0.05 \% & 98.41 $\pm$ 0.05 \% & 98.89 $\pm$ 0.04 \%\\

    & \emph{$P_{star}$} & 98.10 $\pm$ 0.05 \% & 98.41 $\pm$ 0.05 \% & 98.89 $\pm$ 0.04 \%\\

    & \emph{$R_{star}$} & 98.18 $\pm$ 0.05 \% & 98.29 $\pm$ 0.05 \% & 98.90 $\pm$ 0.04 \%\\
    \hline

    \textbf{variability\ features}
    
    & \emph{$P_{quasar}$} & 95.73 $\pm$ 0.07\% & 97.13 $\pm$ 0.07 \% & 98.54 $\pm$ 0.04 \% \\
    
    & \emph{$R_{quasar}$} & 94.87 $\pm$ 0.09\% & 96.16 $\pm$ 0.08 \% & 97.74 $\pm$ 0.05 \% \\
      
    & \emph{$P_{star}$} & 94.92 $\pm$ 0.08\% & 96.20 $\pm$ 0.08 \% & 97.76 $\pm$ 0.05 \% \\
          
    & \emph{$R_{star}$} & 95.73 $\pm$ 0.08\% & 97.15 $\pm$ 0.07 \% & 98.55 $\pm$ 0.04 \% \\
    \hline
   
    \textbf{color\ features}
    
    & \emph{$P_{quasar}$} & --- & --- & 97.08 $\pm$ 0.06\% \\
    
    & \emph{$R_{quasar}$} & --- & --- & 96.68 $\pm$ 0.08\% \\
      
    & \emph{$P_{star}$} & --- & --- & 96.70 $\pm$ 0.07\% \\
         
    & \emph{$R_{star}$} & --- & --- & 97.09 $\pm$ 0.06\% \\
    \hline
    \hline
    
    \end{tabular}
    \caption{The performance (precision and recall, similar to Table. \ref{t_result}) of three trained RFC models using one-year, two-year and ten-year long light curves.  
    }
   \label{t_onetwo}
\end{table*}

\begin{figure}[htbp]
\includegraphics[width=1\linewidth]{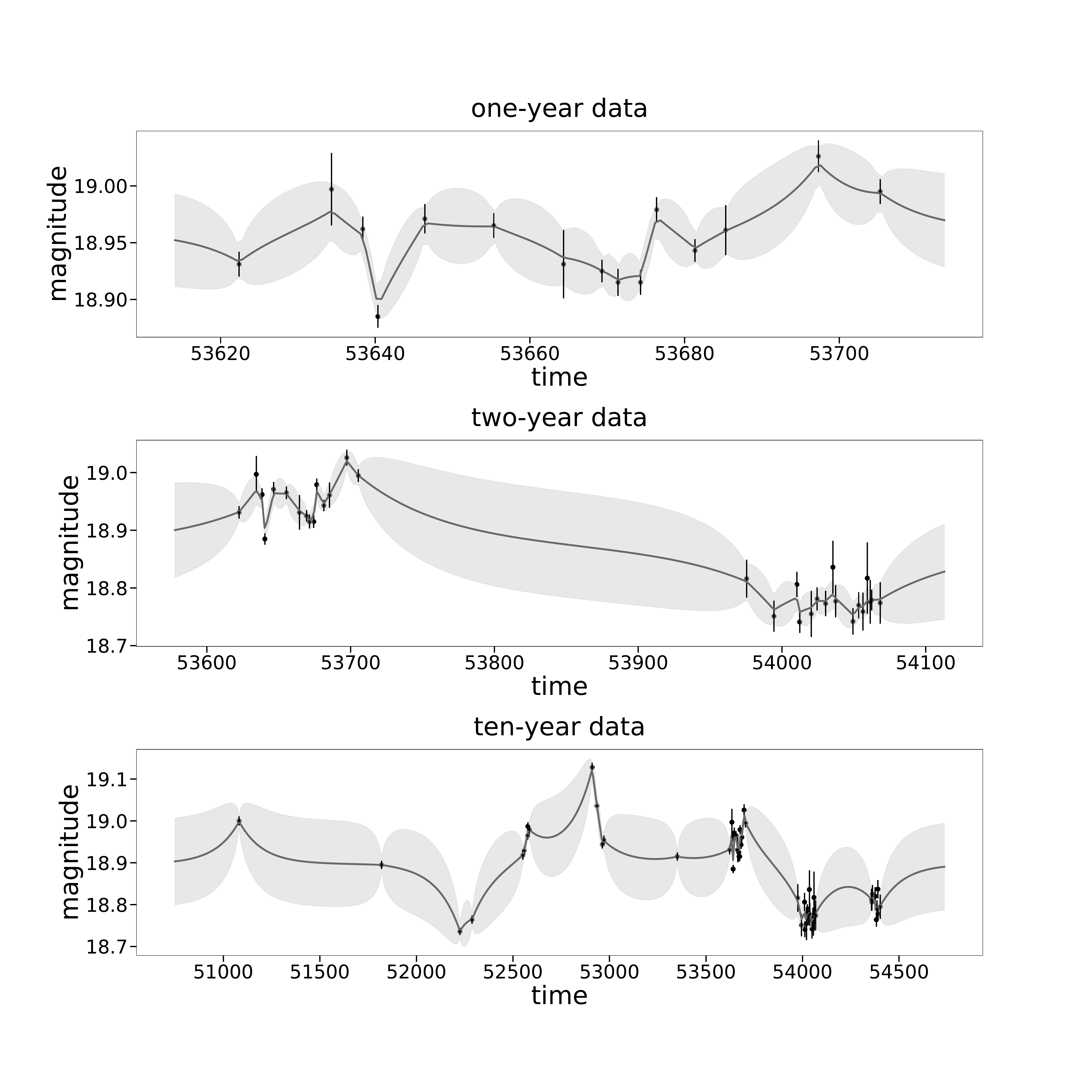}
\caption{From top to bottom: full ten-year, two-year (2005-2006), and one-year (2005) SDSS g band light curves of an example quasar, and the corresponding best-fitted DRW models.}
\label{oneyear}
\end{figure}

In future surveys, pre-selection of quasar candidates could be required when only one or two semesters of time domain observations are available.
Below we explore whether the performance of the variability-based quasar selection is sensitive to the length of the light curves utilized to derived the variability parameters.
This is realized through feeding one-year and two-year data extracted from the Stripe 82 light curves to the classifiers. The one-year data are specified as data collected in 2005, and two-year as data collected in 2005 \& 2006. The typical  number of epochs in one-year and two-year data are 15 and 30 respectively.
Example quasar light curves are given in Fig. \ref{oneyear}.

Following the procedures described in \S\ref{s_ml}, we train three RFC models using ten-year, two-year and one-year datasets respectively. For all models we use samples consisted of randomly selected 2800 quasars plus 2800 stars\footnote{Here we use 2800:2800 sources (instead of 3000:3000) to train the classifiers as for some stellar sources the DRW fitting fails for one-year or two-year datasets.} to train, and 600 quasars plus 600 stars to test. With 100 trials we present the derived confusion matrixes with mean precision and recall in Table \ref{t_onetwo}.

\begin{figure*}[htbp]
    \centering 
    \includegraphics[width=1\linewidth]{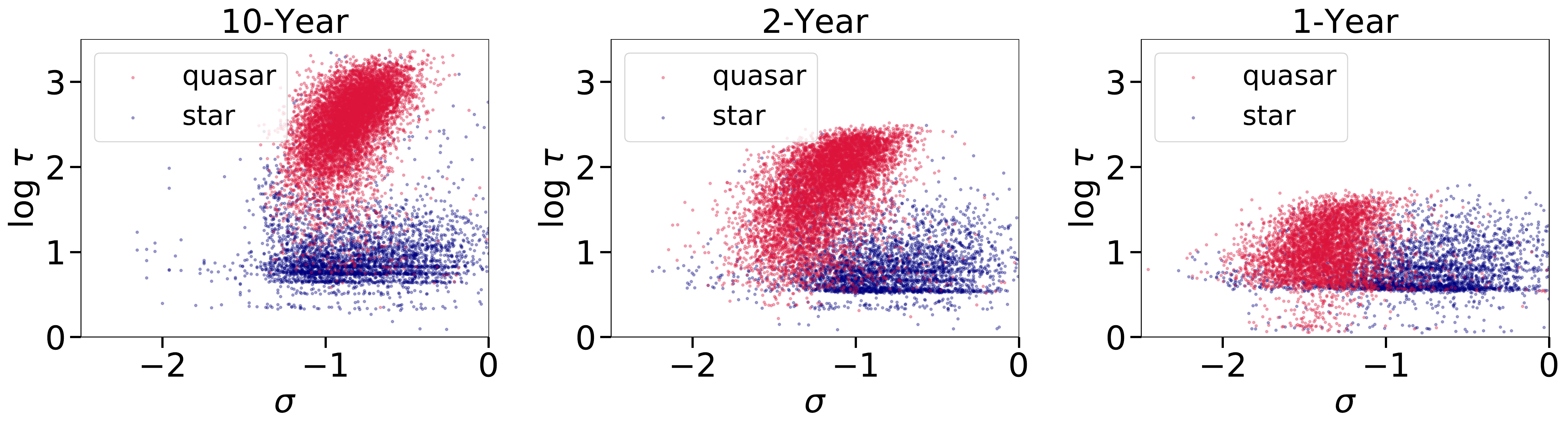}
\caption{$\tau$ vs $\sigma$ for different datasets. Left to right: DRW parameters derived from ten-year, two-year and one-year g band light curves, respectively.
A similar version of the left panel could be see in \cite{2011ApJ...728...26M}.}
    \label{f_idk}
\end{figure*}

We find that when using shorter light curves, the derived test scores are lower, but only slightly. The lower scores are mainly because shorter light curves yielded smaller $\tau$ for quasars
(see Fig. \ref{f_idk}), making them harder to be distinguished from stars.
Nevertheless, the performance of quasar selection by variability alone using two-year light curves is  similar to that using color features alone. The one-year datasets yield only slightly worse scores, 
demonstrating that selecting quasars by variability is still efficient even when only one semester time domain observations are available (with $\sim$ 13 epochs for stars and $\sim$ 18 epochs for quasars).

\begin{table*}[t]
    \center
    \begin{tabular}{c | c c c c}
    \hline
    \hline
   \emph{} &\textbf{5 epochs} &\textbf{9 epochs} &\textbf{15 epochs} &\textbf{all epochs}\\
    
    \hline
    
    \emph{$P_{quasar}$} & 90.55 $\pm$ 0.15 \% & 92.54 $\pm$ 0.14 \% & 93.74 $\pm$ 0.12 \% & 94.51 $\pm$ 0.13 \% \\

    \emph{$R_{quasar}$} & 92.10 $\pm$ 0.17 \% & 93.71 $\pm$ 0.14 \% & 94.91 $\pm$ 0.13 \% & 95.54 $\pm$ 0.13 \% \\

    \emph{$P_{star}$} & 91.98 $\pm$ 0.16 \% & 93.65 $\pm$ 0.14 \% & 94.86 $\pm$ 0.13 \% & 95.51 $\pm$ 0.13 \% \\

    \emph{$R_{star}$} & 90.36 $\pm$ 0.22 \% & 92.43 $\pm$ 0.15 \% & 93.65 $\pm$ 0.13 \% & 94.43 $\pm$ 0.13 \% \\
    \hline
    \hline
    
    \end{tabular}
    \caption{The performance of RFC using data with different length (variability features alone). From left to right are: 5 epochs, 9 epochs, 15 epochs (all random selected) and all epochs in 2005.}
   \label{t_epoch}
\end{table*}

Finally, we investigate the dependence of the performance on the number of epochs obtained within one observing semester. We select sources which have at least 15 epochs in 2005. 6,836 quasars and 1,624 stars are selected. We then randomly select 5, 9, 15 epochs in 2005 from each source to fit with a DRW. Using variability features alone, we use 1,300 quasars and stars to train and 300 to test. The results averaged after 100 trials are listed in Table \ref{t_epoch}. The results of ``all epochs" are comparable to those of ``one-year variability features" in Table \ref{t_onetwo}, but with higher $R_{quasar}$, $P_{star}$ and lower $P_{quasar}$, $R_{star}$. This is because though the new subsamples (with at least 15 epochs from each source) are smaller than the ``one-year" samples used in \ref{t_onetwo} that we can use only 1300 quasars plus 1300 stars to train (instead of 2800:2800), the new subsamples have on average more epochs from each source from those ``one-year" samples used in Table \ref{t_onetwo}. 
Clearly from Table \ref{t_epoch} we can see that the performance decreases with decreasing number of epochs. Note using even ``5 epochs" within one semester could still reach considerably high precision and recall ($\sim$ $90\%$), further demonstrating the efficiency of variation-based quasar selection.

\section{Conclusions}\label{s_conclusion}

In this work we extensively study variability-based quasar selection through training and testing three data-driven classifiers (random forest, AdaBoost, GBDT) with the 10-year long multi-epoch optical photometric data in SDSS Stripe 82. We fit the SDSS Stripe 82 light curves of spectroscopically confirmed quasars and stars with the DRW process using JAVELIN. The main results of this work include:

1. Trained with the variability features alone, all three models can select quasars with similarly and remarkably high precision and completeness ($\sim$ 98.5\% and 97.5\%, trained and tested with 1:1 samples of quasars and stars), even better than using SDSS colors alone ($\sim$ 97.2\% and 96.5\%). 

2. Combining both variability and color features, we achieve precision and completeness both $\sim$ 99.0\%, consistent with previous similar studies. 

3. Using the trained models, we classify the unlabelled variable sources in Stripe 82, and estimate the completeness  of the spectroscopically identified quasar sample 
in Stripe 82 variable source catalog to be $\sim$ 95\% (for $m_i<19.0$).

4. We present the relative importance of the observational features utilized to classify quasars. The top three most important features are $\tau_g$, $\tau_r$, and $u-g$.

5. We show that variability-based quasar selection could still be highly efficient even when only one- or two-year time domain observations are available.

We also discuss the effects of imbalanced samples used to train and test the classifiers. 

This work is supported by National Science Foundation of China (grants No. 11421303 \& 11890693), National Basic Research Program of China (973 program, grant No. 2015CB857005) and CAS Frontier Science Key Research Program (QYZDJ-SSW-SLH006).

\appendix
\section{Peculiar Fitted DRW Parameters}\label{a_star}
\begin{figure*}[htbp]
\centering
\includegraphics[width=1\linewidth]{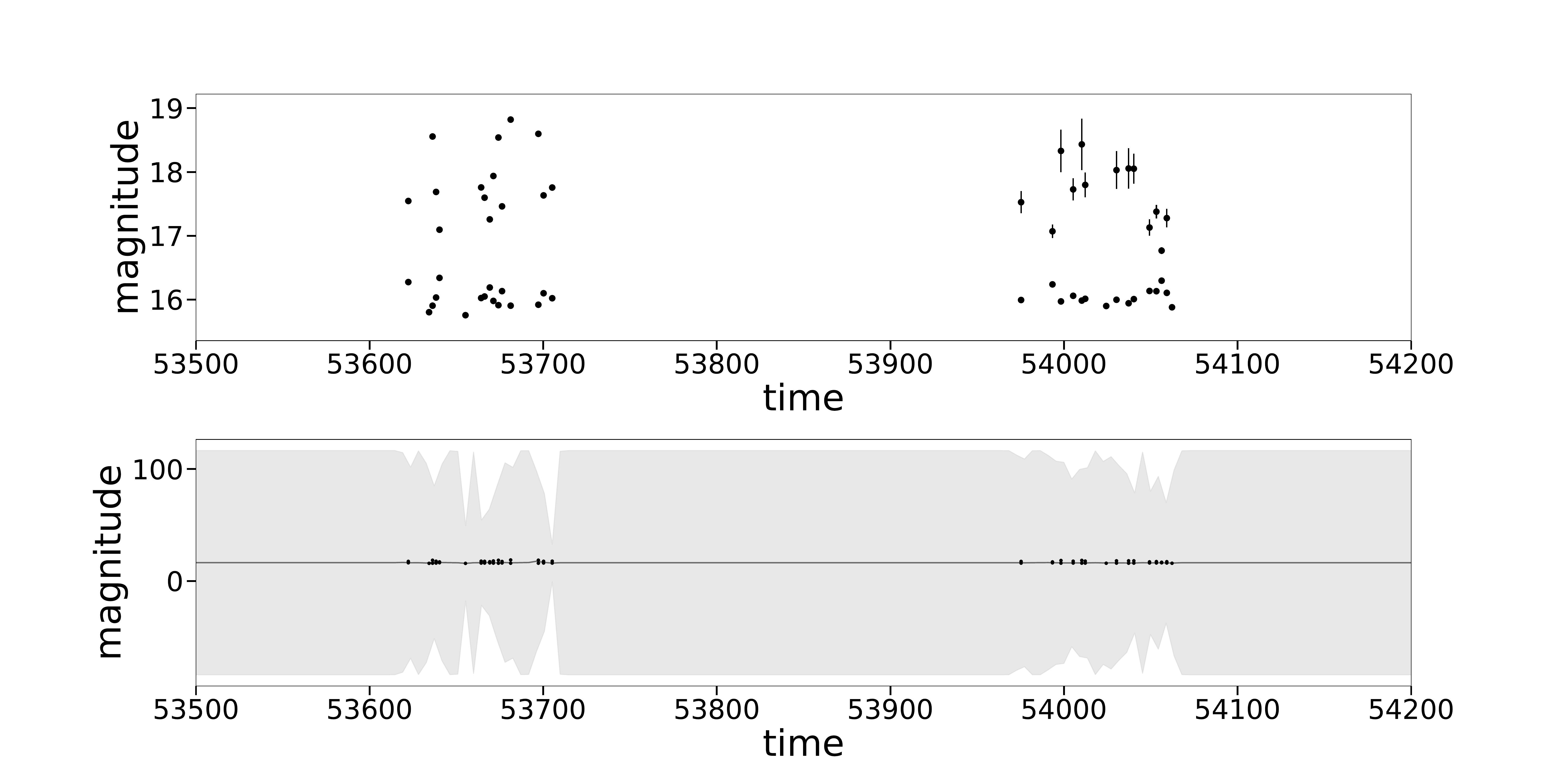}
\caption{Top: An example SDSS g band light curve of a star with extremely large fitted $\sigma$. Bottom: best-fitted DRW model from JAVELIN. Only observations between 53500-54200 are shown in both panels.}
\label{f_star}
\end{figure*}

In Fig. \ref{f_star}, we show an example stellar source whose fitted $\sigma$ is extremely large. From the light curves, we presume that these kind of sources could be cataclysmic variable stars, dwarf novae, etc. Due to the strong variation at very short time scales, the fitted damped random walk parameters could be abnormal, and because of the same reason, they are easy to distinguish.

\clearpage
\bibliographystyle{raa}
\bibliography{paper}

\end{document}